\documentclass[aps,prb,twocolumn,groupedaddress,showpacs]{revtex4}

\usepackage{graphicx}
\usepackage{dcolumn}
\usepackage{bm}
\usepackage{hyperref}
\usepackage[mathlines]{lineno}

\usepackage{color}
\usepackage{amsmath}
\usepackage{amssymb}
\usepackage{epsfig}
\usepackage{wasysym}
\usepackage{bbm}
\usepackage{multirow}

\begin{document}

\title{Prediction of phonon-mediated high temperature superconductivity in stoichiometric Li$_2$B$_3$C}

\author{Miao Gao$^{1}$}

\author{Zhong-Yi Lu$^{1}$}\email{zlu@ruc.edu.cn}

\author{Tao Xiang$^2$}\email{txiang@aphy.iphy.ac.cn}

\affiliation{$^{1}$Department of Physics, Renmin University of
China, Beijing 100872, China}

\affiliation{$^{2}$Institute of Physics, Chinese Academy of
Sciences, P.O. Box 603, Beijing 100190, China }

\begin{abstract}
The discovery of superconductivity in Magnesium Diborate (MgB$_2$) has stimulated great interest in the search of new superconductors with similar lattice structures.
Unlike cuprate or iron-based superconductors, MgB$_2$ is indisputably a phonon-mediated high temperature superconductor.
The emergence of high temperature superconductivity in this material results from the strong coupling between the boron $\sigma$-bonding electrons around the Fermi level and the bond-stretching optical phonon modes.
Here we show, based on the first-principles calculations, that Li$_2$B$_3$C is such a good candidate of superconductor whose superconducting transition temperature (T$_c$) might be even higher than MgB$_2$.
Li$_2$B$_3$C consists of alternating graphene-like boron-carbon layers and boron-boron layers with intercalated lithium atoms between them.
Similar to MgB$_2$, Li$_2$B$_3$C is inherently metallic and possesses two $\sigma$- and two $\pi$-electron bands around the Fermi energy.
The superconducting pairs are glued predominately by the strong interaction between boron $\sigma$-bonding electrons and various optical phonon modes.
\end{abstract}

\date{\today}


\maketitle

A $\sigma$-bond in graphite or other materials is a strong covalent bond of a spin singlet pair formed by two electrons with opposite spins.
In most of materials, a $\sigma$-bond is a stable bound state whose energy sinks well below the Fermi level, making no contribution to the conductivity.
However, if the energy of a $\sigma$-bond is lifted up to the Fermi level by hole doping or other effects, the binding force of the bond is released and electrons become itinerant.
The characteristic feature of the $\sigma$-bond can nevertheless be retained if its hybridization with other conducting electrons is small.
Since the coupling between $\sigma$-electrons and lattice vibrations is generally very large, this may generate a strong attractive interaction to pair electrons to a high-T$_c$ superconducting state.
This is just the physics underlying the 39 K superconductivity in MgB$_2$ discovered by Akimitsu and coworkers in 2000\cite{MgB2}.
This picture was supported by the density functional theory and lattice dynamics calculation, which shows that the electron-phonon interaction between the $\sigma$ bands and bond-stretching optical phonon modes \cite{An-PRL86_4366,Y.Kong-PRB64_020501,Yildirim-PRL87_037001,Choi-PRB66_020513,Choi-Nature418_758} plays a central role in the superconducting pairing in MgB$_2$, and was further confirmed by the isotope effect measurement \cite{Budko-PRL86_1877}.

Stimulated by the discovery of superconductivity in MgB$_2$, great effort has been made to find new phonon-mediated Bardeen-Cooper-Schrieffer (BCS) superconductors during the past decade, focused on chemically substituted MgB$_2$ \cite{Bohnen-PRL86_5771,Mehl-PRB64_140509,Choi-PRB80_064503} or layered compounds similar to MgB$_2$ \cite{Rosner-PRL88_127001,Dewhurst-PRB68_020504}.
Among them the hole-doped LiBC with 50\% Li deficiency has attracted particular interest.
LiBC is isostructural and isovalent to MgB$_2$, with alternatively stacked graphene-like boron-carbon layers intercalated between Li layers \cite{Worle-ZAAC621_1153}.
Without doping, LiBC is a large gap semiconductor with a boron-carbon $\sigma$ band at
the top of valance band\cite{Rosner-PRL88_127001, Worle-ZAAC621_1153,Pronin-PRB67_132502,Karimov-JPCM16_5137}.
If this $\sigma$ band can be rigidly lifted up to cross the Fermi level by hole doping, it was found by the first principle calculation that the electron-phonon coupling constant $\lambda$, which again contributes predominantly by the coupling between the $\sigma$-electrons and the boron-carbon bond-stretching modes, is about 1.8 times larger than in MgB$_2$.
Based on this calculation, it was predicted that Li$_{0.5}$BC could be a high-T$_c$ superconductor with a transition temperature about 100 K.\cite{Rosner-PRL88_127001}
However, the superconductivity has never been observed in hole-doped Li$_x$BC\cite{Bharathi-SSC124_423,Souptela-SSC125_17,Fogg-PRB67_245106,Fogg-CC12_1348}.
The reason is that the hole doping by creating Li deficiency introduces strong lattice distortion to both Li and boron-carbon layers, which changes completely the band structure of LiBC and diminishes the wish to lift the $\sigma$-band up to the Fermi level.\cite{Fogg-JACS128_10043}
Li$_{0.5}$BC remains in a semiconducting phase.

\begin{figure}[th]
\begin{center}
\includegraphics[width=8.6cm]{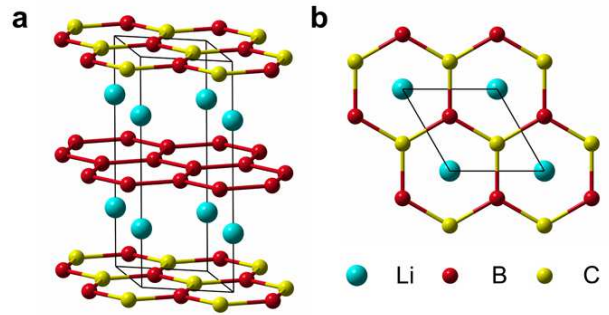}
\caption{ {\bf Crystal structure of Li$_2$B$_3$C}. ({\bf a}) A bird's-eye view. ({\bf b}) Top view. The solid black line denotes the unit cell.}
\label{fig:Structure}
\end{center}
\end{figure}

In this work, instead of doping holes by introducing Li vacancies, we substitute half of the boron-carbon layers by the boron-boron layers in LiBC.
This leads to a new compound, Li$_2$B$_3$C, whose crystal structure is shown in Fig.~\ref{fig:Structure}.
Based on the first-principle density functional theory calculation, we find that Li$_2$B$_3$C is inherently a metal with two $\sigma$-bands across the Fermi surface.
As expected, the coupling of these $\sigma$-bands with optical phonons is very strong, which can drive this material to a high-T$_c$ superconducting state.

The lattice parameters after full relaxation for Li$_2$B$_3$C are shown and compared with the corresponding data for LiBC in Table~\ref{tab:parameter}.
Our theoretical lattice parameters for LiBC agree accurately with the experimental ones along both $a$ and $c$ axes.
Since the radius of boron atom is larger than that of carbon, the lattice parameters of Li$_2$B$_3$C are slightly enlarged in comparison with LiBC.
Li atoms in Li$_2$B$_3$C move slightly towards the honeycomb boron-boron layers, due to the lack of inversion symmetry between boron-carbon and boron-boron layers.

\begin{table}[tbh]
\caption{\label{tab:parameter}
Lattice parameters and fractional coordinates of Li for LiBC and Li$_2$B$_3$C.}
\begin{ruledtabular}
\begin{tabular}{lccc}
    & LiBC & LiBC$^{\text{Expt.}}$ (Ref.\onlinecite{Worle-ZAAC621_1153}) &Li$_2$B$_3$C  \\
\colrule
$a$ (\AA)  & 2.743 & 2.752 & 2.839  \\
$c$ (\AA)  & 6.987 & 7.058 & 7.081 \\
$c/a$ (\AA)  & 2.556 & 2.565 & 2.494 \\
$z_{\text{Li}}$  & 0.25/0.75 & 0.25/0.75 & 0.2547/0.7453  \\
\end{tabular}
\end{ruledtabular}
\end{table}

\begin{figure}[tbh]
\begin{center}
\includegraphics[width=8.6cm]{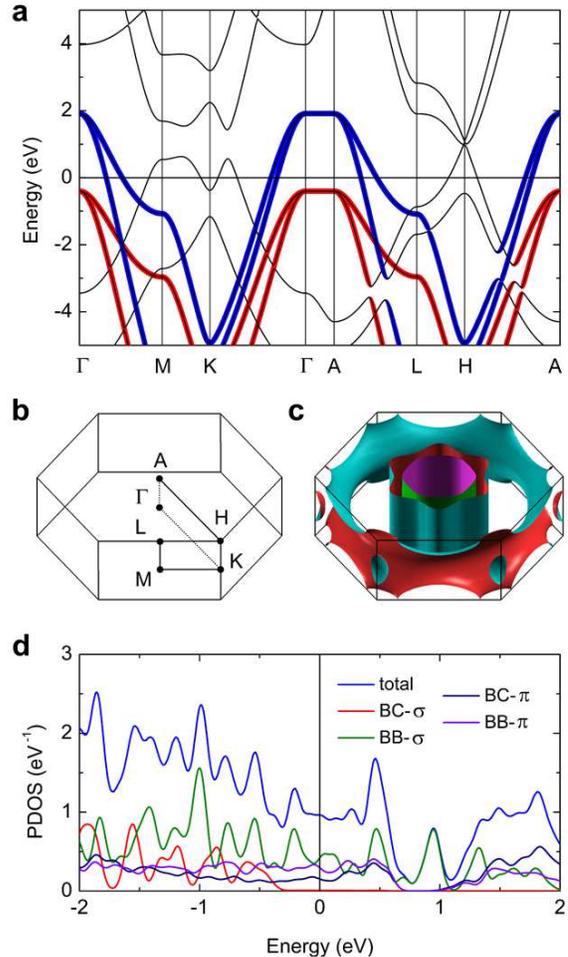}
\caption{ {\bf Electronic structure of Li$_2$B$_3$C}.
(a) Band structure, where the thickness of blue and red line stand for
the weights of the $\sigma$ bonds from the boron-carbon and pure boron layers, respectively. (b) High symmetrical points in Brillouin zone
in fractional coordinates, M $(\frac{1}{2}, 0, 0)$, K $(\frac{1}{3} , \frac{1}{3} ,0)$, A $(0,0,\frac{1}{2})$, L $(\frac{1}{2},0,\frac{1}{2})$, H $(\frac{1}{3}, \frac{1}{3} , \frac{1}{2})$.
(c) The Fermi surfaces. ({\bf d}) Orbital-resolved partial density of states (PDOS). BB and BC represent pure boron and boron-carbon layers, respectively.}
\label{fig:Band}
\end{center}
\end{figure}

Figure~\ref{fig:Band} shows the electronic structure for Li$_2$B$_3$C.
In both boron-carbon and pure boron layers, hybridization of atomic orbitals generate two different kinds of bonds, one is the $sp^2$ hybridized $\sigma$ bond, and the other is the $p_z$ overlapped $\pi$ bond whose energy is generally higher than the former one.
There are four $\sigma$ bands around the Fermi level.
These bands, represented by the blue and red lines in Fig.~\ref{fig:Band}(a), are highly two-dimensional and characterized by the dispersionless energy-momentum curves along the $k_z$ (i.e. $\Gamma$-$A$) direction.
Two of them cross the Fermi level and each exhibits a cylindrical-like Fermi surface around the zone center (Fig.~\ref{fig:Band}(c)).
These two bands are formed mainly by the $\sigma$-electrons in the honeycomb boron layers.
The other two $\sigma$-bands, which are about 0.4 eV below the Fermi energy, are mainly the contribution of $\sigma$-electrons in the boron-carbon layers and confirmed by the result of partial density of states shown in Fig.~\ref{fig:Band}(d).
In comparison with the boron-carbon layer, the pure boron layer has one less electron in every unit cell and can be regarded as an intrinsically hole-doped subsystem.
That is why only the $\sigma$-band from the pure boron layer can appear on the Fermi level.

Besides the two $\sigma$ bands, there are also two $\pi$ bands crossing the Fermi energy.
They show strong dispersion along the $k_z$ direction.
One of them shows a warped hole Fermi surface sheet around the zone boundaries and a Dirac cone-like band dispersion at $H$ and equivalent points about 1 eV above the Fermi level.
The other $\pi$-band shows six hole pocket-like Fermi surface sheets at the zone corners (Fig~\ref{fig:Band}(c)).
Similar as in MgB$_2$, the energy overlap between $\sigma$ and $\pi$ bands at the Fermi level is caused by the attraction of cations.\cite{An-PRL86_4366,Kortus-PRL86_4656}
The tempestuous decrease in the density of states between 0.5 eV and 1.3 eV (Fig.~\ref{fig:Band}(d)) originates in the linear dispersion of Dirac cone-like bands.
In comparison with the $\sigma$ and $\pi$ electrons, the contribution to the density of states near the Fermi level from Li-$2s$ and Li-$2p$ is negligibly small.

\begin{widetext}

\begin{figure}[tbh]
\begin{center}
\includegraphics[width=16cm]{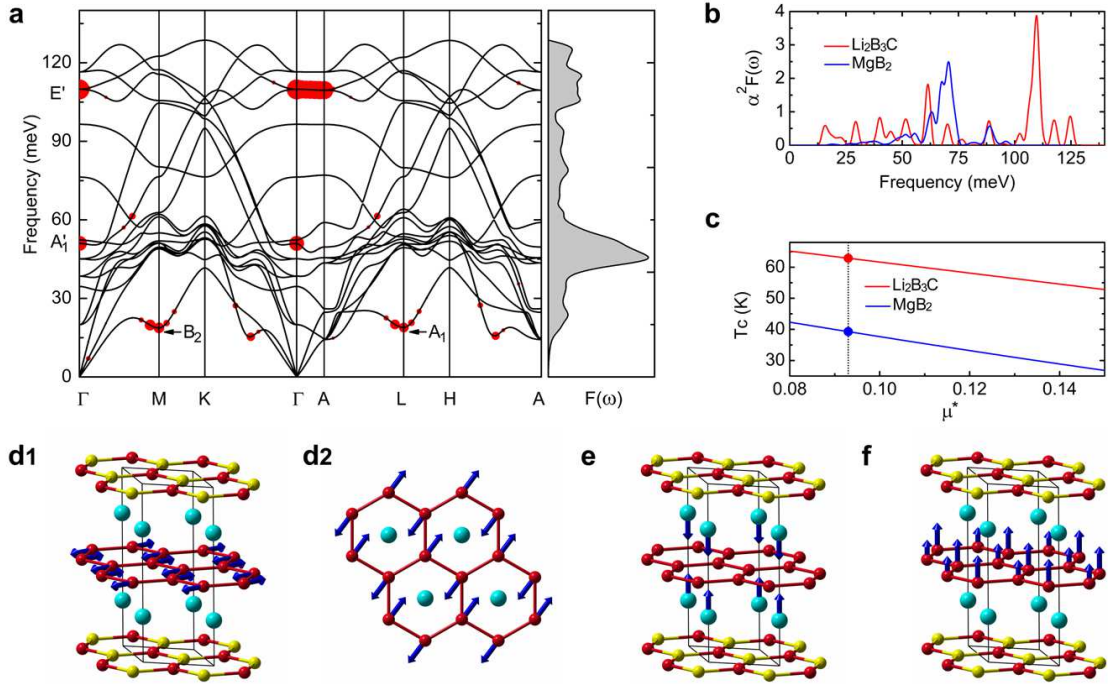}
\caption{
{\bf Calculated lattice dynamics and electron-phonon coupling of Li$_2$B$_3$C}.
({\bf a}) Phonon dispersion curves (left panel) and phonon density of states $F(\omega)$ (right panel). The radius of red circles donates the strength of electron-phonon coupling constant for a given wave vector and mode $\lambda_{{\bf q}\nu}$. $A'_1$, $B_2$, $A_1$, and $E'$ are the phonon modes at high-symmetry points, which have relatively large electron-phonon coupling.
({\bf b}) Eliashberg spectral function $\alpha^2F(\omega)$.
({\bf c}) Evaluated superconducting transition temperature as
a function of $\mu^*$.
({\bf d-f}) Vibrational configurations for the twofold degenerated $E'$ mode at
$\Gamma$ with $\omega\simeq 110$ meV, the $A'_1$ mode at $\Gamma$ with $\omega\simeq 50$ meV, the $B_2$ mode at $M$ or the $A_1$ mode at $L$ with $\omega\simeq 20$ meV, respectively.
The blue arrows represent the directions of vibrations, and their lengthes donate the relative amplitudes.
({\bf d2}) Top view of the $E'$ mode.
}
\label{fig:phonon}
\end{center}
\end{figure}

\end{widetext}

To quantify the electron-phonon interaction, we have also calculated the phonon spectra and the electron-phonon coupling matrix elements.
Fig.~\ref{fig:phonon}(a) shows the phonon dispersion curves along high-symmetry lines in the momentum space and the phonon density of states.
In comparison with the $E_{2g}$ mode in MgB$_2$, we find that there are basically four phonon modes, which contribute most to the electron-phonon coupling in Li$_2$B$_3$C.
They are the $A'_1$ modes at $\Gamma$, the $E'$ mode along the $\Gamma$-$A$ line, the $B_2$ mode at $M$, and the $A_1$ mode at $L$.
A graphical representation of these modes is shown in Fig.~\ref{fig:phonon}(d-f).
$E'$ is a twofold degenerated bond-stretching optical mode, with intra layer movements of boron atoms in the boron layer, which is analogous to $E_{2g}$ mode in MgB$_2$.
$A'_1$ represents an opposite displacement between two Li layers towards the central honeycomb boron layer.
$B_2$ and $A_1$ are two interlayer vibrations of boron atoms in the boron layer.
The electron-phonon coupling constants at the high-symmetry points for these phonon modes are proportional to the radius of red circles shown in Fig.~\ref{fig:phonon}(a).

From the above results, we calculate the Eliashberg spectral function $\alpha^2F(\omega)$ for Li$_2$B$_3$C.
The result is shown in Fig.~\ref{fig:phonon}(b) and compared with that for MgB$_2$.
Our result of the spectral function for MgB$_2$ agrees well with the previous calculations published by Kong {\it et al.}\cite{Y.Kong-PRB64_020501} and by Bohnen {\it et al.}\cite{Bohnen-PRL86_5771}, which have been successfully used to describe the specific heat \cite{Golubov-JPCM14_1353}, tunneling\cite{Brinkman-PRB64_180517}, penetration depth\cite{Golubov-PRB66_054524} and other experimental results\cite{Dolgov-PRB72_024504}.
In MgB$_2$, the spectral function is predominately determined by the coupling  between the $\sigma$-electrons and the bond-stretching $E_{2g}$ mode.
This leads to a single peak dominant structure in $\alpha^2F(\omega)$.
However, for Li$_2$B$_3$C, besides the $E^\prime$ mode, we find that other three modes shown in Fig.~\ref{fig:phonon}, i.e. the $A'_1$, $B_2$ and $A_1$ modes, have also significant contribution to $\alpha^2F(\omega)$.
This is because $\alpha^2F(\omega)$ is determined by the ratio between the phonon linewidth and the frequency.
A moderate phonon linewidth can give a considerable contribution to $\alpha^2F(\omega)$ in the low frequency range.
The phonon linewidth of the $E'$ mode is about sixty times larger than that of $B_2$ and $A_1$ modes.
It gives the highest peak of $\alpha^2F(\omega)$ at 110 meV.
The first peak of $\alpha^2F(\omega)$ at 15 meV in Li$_2$B$_3$C results from the coupling of $\sigma$-electrons with the $B_2$ and $A_1$ modes.

\begin{table}[tbh]
\caption{\label{tab:el-ph}
Parameters of electron-phonon coupling for MgB$_2$ and Li$_2$B$_3$C.
The effective screened Coulomb repulsion constant $\mu^*$
is chosen as 0.093.}
\begin{ruledtabular}
\begin{tabular}{lccc}
 & MgB$_2$ & Li$_{0.5}$BC (Ref.\onlinecite{Rosner-PRL88_127001}) & Li$_2$B$_3$C \\
\colrule
$\lambda$  & 0.85 & 1.5 & 1.53 \\
$\omega_{\text{log}}$ (meV)   & 61.79 & 68 & 45.69 \\
$T_c$ (K)  & 39.3 & 100 & 62.9 \\
\end{tabular}
\end{ruledtabular}
\end{table}

Knowing the Eliashberg spectral function, one can calculate the electron-phonon
coupling constant $\lambda$, the logarithmic average frequency $\omega_{\text{log}}$, and then the superconducting transition temperature $T_c$ from the Allen-Dynes equation\cite{Allen-PRB6_2577,Allen-RPB12_905},
Eq.~(\ref{eq:Tc}).
The results are summarized in Tab.~\ref{tab:el-ph}.
For MgB$_2$, the values of $\lambda$ and $\omega_{\text{log}}$ agree with the results published by other groups\cite{An-PRL86_4366,Y.Kong-PRB64_020501,Bohnen-PRL86_5771,Kortus-PRL86_4656}.
For Li$_2$B$_3$C, the electron-phonon coupling constant $\lambda$ is equal to  1.53, higher than the corresponding value for MgB$_2$ (Table~\ref{tab:el-ph}).
But the value of $\omega_{\text{log}}$ is smaller than that of MgB$_2$ and Li$_{0.5}$BC,
due to the shape of phonon density of states $F(\omega)$.
The value of the effective screened Coulomb potential $\mu^*$ cannot be determined by the first principle calculation.
It generally takes a value  between 0.09 and 0.15.
Choosing $\mu^* = 0.093$ gives T$_c$ = 39.3 K for the reference material MgB$_2$, as observed.
Using the same $\mu^*$, our calculated data of $\lambda$ and $\omega_{log}$ gives T$_c$ = 62.9 K for Li$_2$B$_3$C.
T$_c$ decreases with increasing $\mu^*$.
But even for $\mu^*$ = 0.15, we find that T$_c$ is still above 50 K (Fig.~\ref{fig:phonon}(c)).

Strong electron-phonon coupling of Li$_2$B$_3$C may raise the question of the associated lattice instability. It is likely that Li$_2$B$_3$C is at the edge of some lattice instability due to the unequal distribution of B or C atoms in the two neighboring graphene-like layers. Thus to synthesize this material with conventional chemical or physical synthesis approaches might be difficult. However, considering the fact that the high quality single crystal of LiBC is available or can be easily synthesized in laboratory and the lattice mismatch between LiBC and Li$_2$B$_3$C is very small (see Table \ref{tab:parameter}), we believe that to grow high quality Li$_2$B$_3$C films on a LiBC substrate by the molecular beam epitaxial is highly feasible. The tensile stress between the film and the substrate can also stabilize the crystal structure of Li$_2$B$_3$C even if it has any lattice instability.

In summary, we propose that there is high possibility to find a strong electron-phonon coupled high temperature superconductor if one can lift the $\sigma$-bonding or other strong chemical bonding band to the Fermi surface.
Based on this picture and the first principle calculation, we predict that the stoichiometric Li$_2$B$_3$C is a phonon-mediated high-T$_c$ superconductor with a T$_c$ above 50K.
Strong electron-phonon coupling in this material results from the coupling of the boron $\sigma$-band with the intra-layer bond-stretching phonon mode of boron as well as the inter-layer phonon modes of lithium and boron.

\section*{Methods}

In our density functional theory calculations the plane wave basis method is used \cite{pwscf}.
We adopt the generalized gradient approximation (GGA) with Perdew-Burke-Ernzerhof formula\cite{PBE} for the exchange-correlation potentials.
The ultrasoft pseudopotentials\cite{vanderbilt} are used to model the electron-ion interactions.
After the full convergence test, the kinetic energy cut-off and the charge density cut-off of the plane wave basis are chosen to be 60 Ry and 600 Ry, respectively.
The Gaussian broadening technique of width 0.05 Ry is used.
The self-consistent electron density and the Fermi surface structure are evaluated using a $18\times18\times12$ k-point grid and a $36\times36\times24$ k-point grid, respectively.
Phonon wave vectors are sampled on a $6\times6\times4$ mesh.
The lattice constants after full relaxation, shown in Table~\ref{tab:parameter} are adopted.

The phonon spectra and the electron-phonon coupling constants are calculated based on density functional perturbation theory \cite{Baroni-RMP73_515} and Eliashberg equations \cite{Eliashberg-SPJ11_696,Bergmann-ZP263_59}.
The electron-phonon coupling matrix element $g_{{\bf k},{\bf q}\nu}^{ij}$, is determined by \cite{Allen-PRB6_2577,Allen-RPB12_905}
\begin{equation}
g_{{\bf k},{\bf q}\nu}^{ij}=\Big(\frac{\hbar}{2M\omega_{{\bf q}\nu}}\Big)^{1/2}\langle\psi_{i,{\bf k}}|\frac{dV_{\text{SCF}}}{d\hat{u}_{{\bf q} \nu}}\cdot\hat{e}_{{\bf q} \nu}|\psi_{j,{\bf k+q}} \rangle,
\end{equation}
where $M$ is the atomic mass, ${\bf q}$ and ${\bf k}$ are wave vectors,
$ij$ and $\nu$ donate indices of electronic energy bands and phonon modes, respectively.
$\omega_{{\bf q}\nu}$ and $\hat{e}_{{\bf q} \nu}$ stand for the phonon frequency and eigenvector of the $\nu$-th phonon mode at vector ${\bf q}$.
$V_{\text{SCF}}$ is the self-consistent potential.
$\psi_{i,{\bf k}}$ and $\psi_{j,{\bf k+q}}$ are Kohn-Sham orbitals.

The phonon linewidth $\gamma_{{\bf q}\nu}$ is defined by averaging the coupling matrix element over the Fermi surface \cite{Allen-PRB6_2577,Allen-RPB12_905}.
\begin{equation}
\gamma_{{\bf q}\nu}=\frac{2\pi\omega_{{\bf q}\nu}}{\Omega_{\text{BZ}}}\sum_{ij}\int d^3k|g_{{\bf k},{\bf q}\nu}^{ij}|^2\delta(\epsilon_{{\bf q},i}-\epsilon_F)\delta(\epsilon_{{\bf k+q},j}-\epsilon_F),
\end{equation}
where $\epsilon_{{\bf q},i}$ and $\epsilon_{{\bf k+q},j}$ are eigenvalues of Kohn-Sham orbitals at given bands and wave vectors.
The spectral function is then given by\cite{Allen-PRB6_2577,Allen-RPB12_905}
\begin{equation}
\label{eq:spectral}
\alpha^2F(\omega)=\frac{1}{2\pi N(\epsilon_F)}\sum_{{\bf q}\nu}\delta(\omega-\omega_{{\bf q}\nu})\frac{\gamma_{{\bf q}\nu}}{\hbar\omega_{{\bf q}\nu}},
\end{equation}
where $N(\epsilon_F)$ is the density of states at the Fermi level.

The electron-phonon coupling constant $\lambda$ is determined by integrating the spectral function over the first Brillouin zone\cite{Allen-PRB6_2577,Allen-RPB12_905},
\begin{equation}
\label{eq:lambda}
\lambda=\sum_{{\bf q}\nu}\lambda_{{\bf q}\nu}=2\int\frac{\alpha^2F(\omega)}{\omega}d\omega,
\end{equation}
where
\begin{equation}
\lambda_{{\bf q}\nu}=\frac{\gamma_{{\bf q}\nu}}{\pi\hbar N(e_F)\omega^2_{{\bf q}\nu}}.
\end{equation}

Finally, by utilizing the Allen-Dynes formula\cite{Allen-PRB6_2577,Allen-RPB12_905}
we can find the superconducting transition temperature
\begin{equation}
\label{eq:Tc}
T_c=\frac{\omega_{\text{log}}}{1.2}\exp\Big[\frac{-1.04(1+\lambda)}{\lambda(1-0.62\mu^*)-\mu^*}\Big],
\end{equation}
where $\mu^*$ is an effective screened Coulomb potential which usually takes a value between 0.09 and 0.15\cite{Rosner-PRL88_127001,Sakamoto-JPSJ65_489,Richardson-PRL78_118,Lee-PRB52_1425}, $\omega_{\text{log}}$ is the logarithmic average frequency,
\begin{equation}
\label{eq:omega_log}
\omega_{\text{log}}=\exp\Big[\frac{2}{\lambda}\int\frac{d\omega}{\omega}\alpha^2F(\omega)\log\omega\Big].
\end{equation}

\begin{acknowledgements}
This work is supported by National Natural Science Foundation of China (Grant No. 11190024) and National Program for Basic Research of MOST of China (Grant No. 2011CBA00112).
\end{acknowledgements}

\end{document}